\documentclass[pra,aps,10pt,superscriptaddress,twocolumn]{revtex4-1}
\usepackage{graphicx}\usepackage[colorlinks=true,linkcolor=blue,citecolor=red,urlcolor=magenta]{hyperref}
\usepackage{amsmath}
\usepackage{color}

\begin{document}

\title{Signatures of quantum chaos in low-energy mixtures of few fermions}

\author{Patrycja  \L yd\.{z}ba}
\affiliation{Department of Theoretical Physics, Wroclaw University of Science and Technology, 50-370 Wroc{\l}aw, Poland}

\author{Tomasz Sowi\'{n}ski}
\affiliation{\mbox{Institute of Physics, Polish Academy of Sciences, Aleja Lotnik\'{o}w 32/46, PL-02668, Warsaw, Poland}}
\affiliation{\mbox{Instituto Universitario de Matem\'atica Pura y Aplicada, Universitat Polit\`ecnica de Val\`encia, ES-46022 Val\`encia, Spain}}

\begin{abstract}
The low energy dynamics of mesoscopic systems strongly depends on the presence of internal equilibration. For this reason, a better interpretation of ultracold atom experiments requires a more accurate understanding of how quantum chaos manifests itself in these systems. In this paper, we consider a simple but experimentally relevant one-dimensional system of a few ultracold fermions moving in a double-well potential. We analyze its many-body spectral properties, which are commonly used to trace quantum chaos. We observe some signatures of quantum chaos already in the system with three particles. Generally, these signatures become more pronounced when fermions are evenly added to both components. On the contrary, they become suppressed when the particle imbalance is increased.
\end{abstract}

\maketitle

\section{Introduction}
Spectral and dynamical peculiarities of isolated quantum systems have been of interest for a long time. Early studies focused mainly on single-particle systems having few degrees of freedom. Prominent examples are billiards confined by various types of boundaries~\cite{Bohigas_1984,Bohigas_1984b,Aurich_1994,Stockmann_1999} or a kicked rotor with a varied strength of periodical kicks~\cite{Casati_1979,Chirikov_1979}. These and following studies demonstrated that the so-called quantum chaotic models, which display the chaotic dynamics in the semi-classical limit, have universal properties of energy eigenvalues~\cite{Montambaux_1993,Prosen_1999,Rabson_2004,Kolovsky_2004,Kollath_2010} and eigenvectors~\cite{Page_1993,Deutsch_2013,Beugeling_2015,Vidmar_2017,Garrison_2018}. The universal properties are consistent with the properties of random matrices drawn from the Gaussian Orthogonal Ensemble (GOE), provided that the time reversal symmetry is present~\cite{Poilblanc_1993}. Among others, the energy eigenvalues are correlated and avoid crossings~\cite{MEHTA_1967,Haake_2010}, the distribution of spacings is given by the Wigner-Dyson distribution~\cite{Wigner_1957}, while the spectral form factor has a linear ramp~\cite{Cotler_2017,Liu_2018,Chen_2018}. Usually, the aforementioned attributes are observed far from the tails of the spectrum. On the contrary, the so-called quantum integrable systems, which have an extensive number of local integrals of motion that confine the dynamics to periodic orbits in the semi-classical limit, share some properties with the Poisson ensemble or are simply non-universal~\cite{Casati_1985,Hsu_1993,GGE_2016}. When the attention shifted from single-particle to many-body systems, the definition of quantum chaos had to be made independent of the dynamical behaviour in the semi-classical limit, which is either very difficult or impossible to determine. It has been replaced by the universal properties of energy eigenvalues and eigenvectors \cite{Deutsch_2018,D_Alessio_2016,Srednicki_1994}.

Quantum chaos is not merely a theoretical concept. It is a necessary ingredient for particular phenomena to occur (like the thermalization of isolated quantum systems driven out of equilibrium~\cite{Rigol_2007,Rigol_2009,Rigol_2011,Khatami_2012,Caux_2013,Reimann_2016,Lydzba_2021}) or not (like the many-body localization in lattice systems with a correlated~\cite{Kotthoff_2021,Schreiber_2015} or random~\cite{Suntajs_2020,Suntajs_2020b,Bauer_2013,Pal_2010,Huse_2007} disorder). This line of research has been further intensified by recent advances in experiments with ultracold atoms. These kinds of setups are almost perfectly isolated from the environment and have highly controllable internal parameters~\cite{Kinoshita_2004,Hofferberth_2007,Trotzky_2008,Trotzky_2012,Edge_2015,Mazurenko_2017,Sowinski_2019}. The external confinements, atomic numbers and inter-atomic interactions can be tuned with a great precision~\cite{Cheng_2010,2011SerwaneScience,2013WenzScience}. The ultracold atom experiments allow adopting the bottom-up approach, in which one witnesses how many-body effects emerge in the system as one gradually increases the particle number. Many papers have addressed this issue from the theoretical point of view, also in the context of quantum chaos~\cite{Fogarty_2021,Mirkin_2021,Santos_2010,Masud_2021}. Nevertheless, there is a room for further research, since the latter studies mainly focus on the high energy limit, which is not always justified for the low temperature experiments.

Recently, ultracold setups with two-component mixtures of few fermions loaded to a one-dimensional double-well potential were realized~\cite{2011SerwaneScience,2013WenzScience}. Furthermore, a full control of their quantum state was achieved, {\it e.g.}, the system could be initialized with an arbitrary configuration of particles in two wells, while tunneling rates and interactions could be independently controlled \cite{Murmann_2015}. Subsequently, a corresponding model was studied numerically~\cite{Sowinski_2016,Nandy_2020}. Typically, the system is considered as prepared in a state with fermions from different components occupying different wells. Therefore, the initial interaction energy is almost negligible. However, when particles tunnel through the barrier, interactions start to play a significant role. For example, the time evolution of the number of particles in wells depends on their strength. This effect was experimentally studied in the fluorescence measurements~\cite{Murmann_2015}. It turned out that in the minimal setup with two opposite-spin fermions, the flow of particles is rather regular, with the transmission rate dependent on the interaction strength. However, the time evolution became highly unpredictable in systems with a larger number of particles. By unpredictability we mean that a small change of the interaction strength led to an entirely different dynamics of the system. It was suggested, but never demonstrated, that this may be explained by the quantum chaotic properties of the many-body spectrum. In our work we aim to verify this hypothesis.

This paper is divided as follows. We first introduce the studied system in Sec.~\ref{sec::sys}, and then the exact diagonalization method, which we implement to establish the low-energy tail of the many-body spectrum in Sec.~\ref{sec::met}. Next, in Sec.~\ref{sec::spe}, we introduce the well-known measures of quantum chaos, like the ratio of level spacings and its distribution. The balanced mixtures with different number of particles are considered in Sec.~\ref{sec::bal}. In the minimal setup with two fermions, we witness no signatures of quantum chaos. In the case of four fermions,  depending on the interaction strength, we observe the coexistence of pseudo-integrable properties with quantum-chaotic properties. Generally, the low-energy tail of the many-body spectrum gradually becomes universal as the number of particles increases. Next, in Sec.~\ref{sec::imb}, we focus on the imbalanced mixtures having a single particle in one component, and a varied number of particles in the other component. We show that quantum chaos emerges more readily when the particle numbers are balanced. Finally, in Sec.~\ref{sec::rol}, we study the effect of the shape of a confining potential on the spectral statistics, and we conclude in Sec.~\ref{sec::con}.

\section{System}\label{sec::sys}
We consider a one-dimensional two-component mixture of few fermions confined in a double-well potential and interacting via contact interactions. The Hamiltonian of this system reads
\begin{equation}
\label{eqH}
\hat{H}  = \int\!\mathrm{d}x \left[ \sum_{\sigma} \hat{\psi}_\sigma^\dagger(x){H}_0\hat{\psi}_\sigma(x) + g\, \hat{n}_\uparrow(x)\hat{n}_\downarrow(x) \right],
\end{equation}
where the fermionic field operator $\hat{\psi}_\sigma(x)$ annihilates a $\sigma$-component particle at $x$ and $\hat{n}_\sigma(x)=\hat{\psi}_\sigma^\dagger(x)\hat{\psi}_\sigma(x)$ is the corresponding density operator. The single-particle Hamiltonian is defined as
\begin{equation}
\label{eqH0}
{H}_{0} = -\frac{\hbar^2}{2m}\frac{\mathrm{d}^2}{\mathrm{d}x^2} + \frac{m\Omega^2}{2} x^2 + \frac{V_0}{\sqrt{\pi}x_0}\mathrm{e}^{-x^2/x_0^2}.
\end{equation}
Note that the mass $m$, the harmonic oscillator frequency $\Omega$, the barrier height $V_0$ and width $x_0$ are the same for both $\sigma$-components. The Hamiltonian~$\left(\ref{eqH}\right)$ commutes with the number operators $\hat{N}_\sigma=\int\!\mathrm{d}x\,\hat{n}_\sigma(x)$. Therefore, we examine its spectral properties in the subspaces with well-defined particle numbers $N_\uparrow$ and $N_\downarrow$. Let us emphasize that the quantum number $\sigma\in\left\{\uparrow,\downarrow\right\}$ is used to distinguish types of atoms trapped in a double-well potential or it is related to a hyperfine degree of freedom controllable by a magnetic field in ultra-cold atom setups~\cite{Weiner_1999}. For further convenience, we express all energies, momenta, and lengths in units of $\hbar\Omega$, $\sqrt{\hbar m \Omega}$, and $\sqrt{\hbar/m\Omega}$, respectively. To make the discussion as clear as possible, we first consider a  particular potential barrier ($V_0=4$ and $x_0=0.2$ in these units), and then briefly discuss the role of its shape on the spectral properties.

The spectrum of the single-particle Hamiltonian~\eqref{eqH0} resembles that of a simple harmonic oscillator in the high energy limit. Therefore, the many-body levels become regular-like (i.e., quasi-equidistant) for both small interactions and large energies. In this manuscript, we consider a quite high barrier height $V_0=4$, and no more than $\mathcal{N}=500$ lowest many-body levels. Therefore, we see its clear footprint in the considered part of the spectrum, even in the noninteracting case (first column in Fig.~\ref{fig1}).

\section{Method}\label{sec::met}
We represent the single-particle Hamiltonian~\eqref{eqH0} as a tri-diagonal matrix on a dense spatial grid, and numerically determine its eigenenergies $\epsilon_i$ and eigenfunctions $\phi_i(x)$. Next, we decompose the field operators $\hat{\psi}_\sigma(x) = \sum_i \hat{a}_{\sigma i} \phi_i(x)$, rewrite the many-body Hamiltonian $\hat{H} = \sum_{\sigma} \sum_{i} \epsilon_i \hat{a}_{\sigma i}^{\dagger} \hat{a}_{\sigma i} + g\sum_{ijkl} U_{ijkl} \hat{a}_{\uparrow i}^{\dagger}\hat{a}^{\dagger}_{\downarrow j} \hat{a}_{\downarrow k}\hat{a}_{\uparrow l}$, and perform its exact diagonalization in a subspace of the Hilbert space spanned by the appropriately selected Fock states. We abandon the conventional orbital cut-off method, in which the many-body basis is a set of Fock states with orbitals $\phi_i(x)$ restricted to $i \le i_\text{max}$. Instead, we employ the energy cut-off method, in which the many-body basis comprises all Fock states with non-interacting energies lower than $E_\text{max}$. The latter method is numerically more efficient, and has been employed in several studies~\cite{1998HaugsetPRA,Chrostowski_2019,2020RojoMathematics}. We select $E_\mathrm{max}\in\left[40,45\right]$, so that the Hilbert subspace is spanned by $778$ Fock states for the smallest system $N_\uparrow=N_\downarrow=1$, while $61848$ for the largest system $N_\downarrow=N_\uparrow=3$. The details of the energy cut-off method and the discussion concerning its accuracy can be found in the Appendix~\ref{Zalacznik}.

It is known that the existence of internal symmetries may lead to additional degeneracies in the many-body spectrum and distorts its properties. In order to witness signatures of quantum chaos, it is thus necessary to limit considerations to the symmetry-invariant subspaces of the Hilbert space. The considered mixture of fermions has a few global symmetries. As already mentioned, the Hamiltonian commutes with the number operators $\hat{N}_\sigma=\int\!\mathrm{d}x\,\hat{n}_\sigma(x)$. We examine its spectral properties in the subspaces with well-defined particle numbers $N_\uparrow$ and $N_\downarrow$. The Hamiltonian~\eqref{eqH} is also invariant under the left-right mirror $Z_2$ symmetry. Hence, we divide its Fock basis into two sectors with different parities $p\in\{+,-\}$ (even and odd sector, respectively). Finally, the Hamiltonian is also SU($2$) symmetric, since it commutes with $\hat{S}^2=\sum_i \hat{S}_i^2$ and $\hat{S}_z$ operators, where $\hat{S}_{i}=\hbar/2 \int \mathrm{d}x\sum_{\alpha\alpha'}\hat{\psi}_\alpha^\dagger(x)\sigma_{\alpha\alpha'}^{i}\hat{\psi}_{\alpha'}(x)$, $\alpha,\alpha'\in\left\{\uparrow,\downarrow\right\}$, $i\in\left\{x,y,z\right\}$, and $\sigma^i$ are three Pauli matrices.  It turns out, however, that lifting this symmetry is not necessary to witness signatures of quantum chaos, as explained in the Appendix \ref{Zalacznik2}. This is appealing, since restricting the evolution of a quantum state to a single spin sector ({\it i.e.}, with a fixed magnitude of the total spin $S$) in the experimental setup is difficult.

The Hamiltonian~\eqref{eqH} is a sparse and block diagonal (due to the parity symmetry) matrix in the Fock basis obtained with the energy cut-off method. Finally, we perform its exact diagonalization and determine the low-energy part of the many-body spectrum $E_i^{(p)}$, which is divided into two parity-invariant sectors $p\in\{+,-\}$.

\section{Spectral statistics}\label{sec::spe}
In our work, we focus on the spectral properties of the Hamiltonian \eqref{eqH} depending on the number of particles $N_\uparrow$ and $N_\downarrow$ as well as the interaction strength $g$. We consider three spectral measures of quantum chaos. They are based on the analysis of the spacings between the nearest energy levels, $\delta_i^{(p)}=E_i^{(p)}-E_{i-1}^{(p)}$, which are independently calculated in parity-invariant sectors $p=\pm$. The first two measures exploit their ratios
\begin{equation}
\tilde{r}_{i}^{(p)}=\text{min}\left(\delta_i^{(p)},\delta_{i+1}^{(p)}\right)/\text{max}\left(\delta_i^{(p)},\delta_{i+1}^{(p)}\right).
\end{equation}
A significant advantage of the ratios over the spacings is that $0 \le\tilde{r}_i^{(p)}\le 1$ for all $i$, so it is not necessary to perform the so-called spectral unfolding to eliminate the influence of the secular part of the density of states~\cite{Oganesyan_2007,Atlas_2013}.

In the first attempt to unveil universal correlations between energy levels, we determine the following average
\begin{equation} \label{aver}
\left<\tilde{r}\right>=\sum_{i=1}^{\mathcal{N}}\left(\tilde{r}_{i}^{(+)}+\tilde{r}_{i}^{(-)}\right)/\left(2\mathcal{N}\right).
\end{equation}
In the above averaging, the sum runs over $\mathcal{N}\in\left\{50,...,500\right\}$ lowest energy states, except for the ground and first excited states that are always well-isolated from the rest of the many-body spectrum. It has been numerically verified that for the Gaussian Orthogonal Ensemble (GOE), {\it i.e.}, an ensemble of real matrices with Gaussian distributed entries, the average $\left<\tilde{r}\right>_\text{GOE} \approx 0.5307$~\cite{Atlas_2013}. Whereas for the Poisson ensemble, {\it i.e.}, uncorrelated energy levels, the average $\left<\tilde{r}\right>_\text{P}\approx 0.38629$~\cite{Atlas_2013}. Thus, it is convenient to introduce the rescaled measure
\begin{equation}
\left<\tilde{r}\right>_\text{n}=\frac{\left<\tilde{r}\right>-\left<\tilde{r}\right>_\text{P}}{\left<\tilde{r}\right>_\text{GOE}-\left<\tilde{r}\right>_\text{P}}
\end{equation}
which interpolates between two extreme cases --  the Poisson distribution typically established in integrable Hamiltonians ($\left<\tilde{r}\right>_\text{n}=0$) and the Wigner-Dyson distribution which is a hallmark of quantum-chaotic Hamiltonians ($\left<\tilde{r}\right>_\text{n}=1$). We use $\left<\tilde{r}\right>_\text{n}$ as the first measure of chaoticity in the considered mixture of fermions.

Instead of considering the average ratio $\langle\tilde{r}\rangle$, one can determine the whole distribution of ratios $P\left(\tilde{r}\right)$ and compare it to the analytical expressions
\begin{subequations}
\begin{align}
P_\text{GOE} \left(\tilde{r}\right)&= \frac{27}{4}\frac{\tilde{r}+\tilde{r}^2}{\left(1+\tilde{r}+\tilde{r}^2\right)^{5/2}},\label{eqGOE} \\
P_\text{P}\left(\tilde{r}\right) &= \frac{2}{\left(1+\tilde{r}\right)^2} \label{eqP}.
\end{align}
\end{subequations}
The above expressions have been obtained for the GOE of $3\times 3$ matrices (which validity has been confirmed for asymptotically large matrices) and for uncorrelated energy levels, respectively~\cite{Atlas_2013}. When calculating this measure of chaoticity, we establish the histograms $P\left(\tilde{r}\right)$ from ratios gathered from both parity-invariant sectors $\tilde{r}_i^{(\pm)}$.

\begin{figure*}
\includegraphics[width=\textwidth]{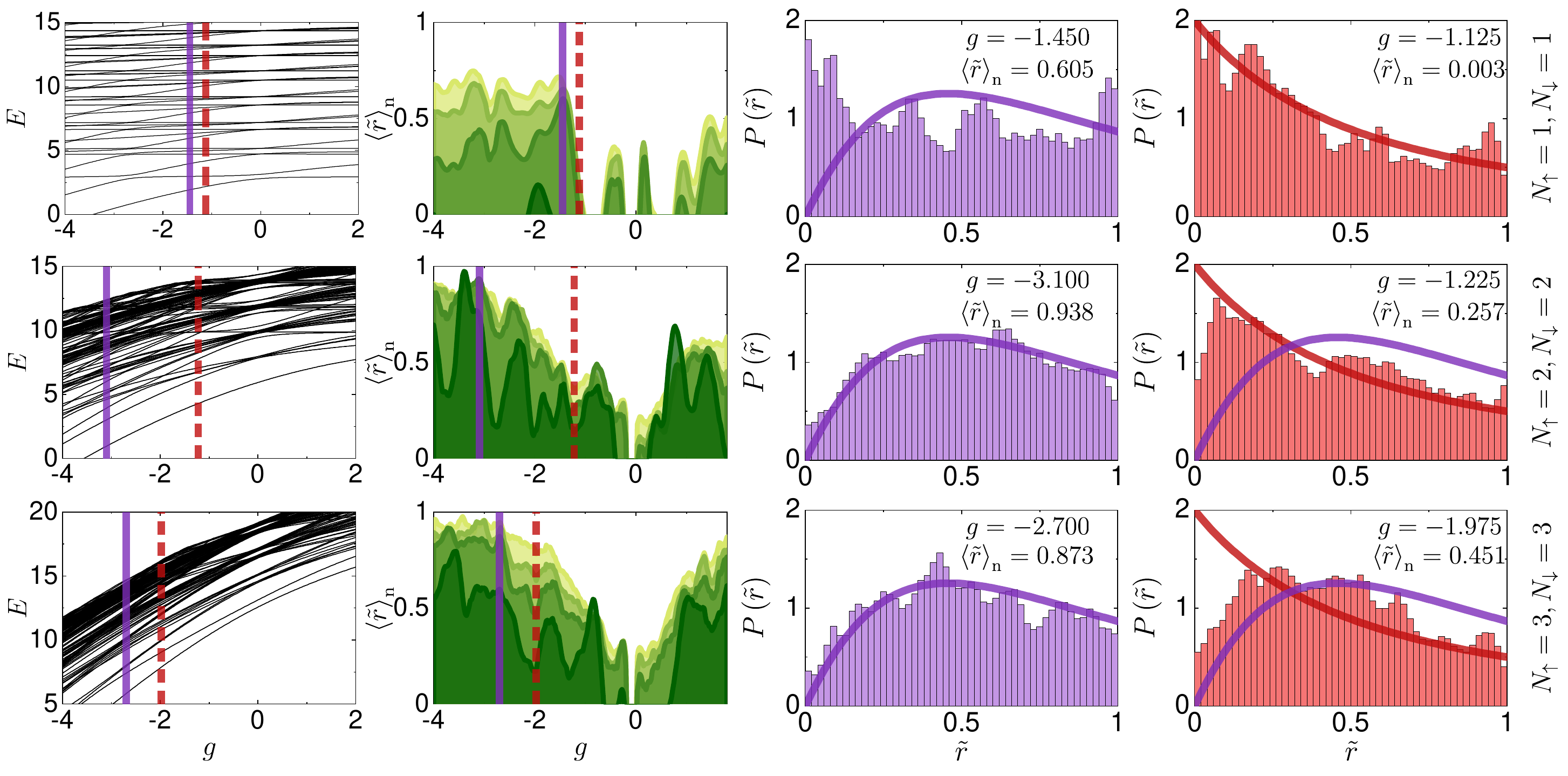}
\vspace{-0.6cm}
\caption{Spectral properties of balanced mixtures with $N_\uparrow=N_\downarrow=1,2,3$ particles. The low-energy part of the many-body spectrum is displayed in the first column. For clarity, from one of the parity-invariant subspaces. The rescaled ratio $\left<\tilde{r}\right>_\text{n}$ is presented in the second column. We consider $\mathcal{N}=50,150,250$ and $500$ ($300$ for $N_\uparrow=N_\downarrow=1$), and numerical results for a  larger $\mathcal{N}$ are marked by a brighter green colour. The histogram $P\left(\tilde{r}\right)$ for a maximum of $\left<\tilde{r}\right>_\text{n}$ and $\mathcal{N}=150$ is presented in the third column [together with the distribution from Eq.~$\left(\ref{eqGOE}\right)$]. Simultaneously, the histogram $P\left(\tilde{r}\right)$ for a minimum of $\left<\tilde{r}\right>_\text{n}$ and $\mathcal{N}=150$ is presented in the fourth column [together with the distribution from Eq.~$\left(\ref{eqP}\right)$]. The interaction strength $g$ corresponding to a maximum (minimum) of $\left<\tilde{r}\right>_\text{n}$ is marked by a purple solid (red dashed) line in the first two columns. }
\label{fig1}
\end{figure*}

In the third attempt, we determine the histogram of spacings ${\cal P}(\delta)$. This requires separating the oscillating part from the secular part of the density of states, so that the mean  density of states is unity. The latter is known as the spectral unfolding~\cite{Hsu_1993,Haake_2010,Morales_2011,Abualenin_2018}. We perform it separately in each of the parity-invariant sectors. We begin with determining a cumulative spectral function ${\bf N}_{p}(E) =\sum_{i=1}^{\mathcal{N}} \vartheta\!\left(E - E_i^{(p)}\right)$ that counts the number of levels with energies lower or equal to $E$ ($\vartheta(.)$ is the Heaviside step function). Then, we separate the global trend $\overline{{\bf N}_{p}(E)}$ by either fitting a high order polynomial with fifteenth degree~\cite{Santos_2010,Morales_2011,Jansen_2019}, or considering linear fits on small intervals around $E_i^{(p)}$ and then the moving average throughout the spectrum~\cite{French_1971}. Both methods provide consistent results for all considered systems, except for the minimal balanced scenario $N_\uparrow=N_\downarrow=1$ that is to some extent sensitive to the unfolding method. Finally, we perform the mapping $E_i^{(p)}\rightarrow\overline{{\bf N}_p\left(E_i^{(p)}\right)}$ and calculate the histogram ${\cal P}(\delta)$ using spacings gathered from both parity-invariant sectors $\delta_i^{(\pm)}$. Next, we compare it with the so-called Brody distribution~\cite{Brody_1981}
\begin{equation} 
\label{BrodyDist}
{\cal{P}}_\gamma(\delta) = \left(\gamma+1\right)b\delta^\gamma\exp\left(-b\delta^{\gamma+1}\right)
\end{equation}
with $\gamma$ standing for the Brody parameter and $b =\Gamma^{\gamma+1}[(\gamma+1)/(\gamma+2)]$ ($\Gamma(.)$ is the Euler's gamma function). Note that the Poisson distribution and the Wigner-Dyson distribution are recovered from the Brody distribution when $\gamma=0$ and $\gamma=1$, respectively. Therefore, the Brody parameter $\gamma$  is a similar measure of chaoticity as $\left<\tilde{r}\right>_\text{n}$. We establish the Brody parameter $\gamma$ from the numerically obtained histograms by the least squares method, in which $\gamma$ is gradually changed from zero to one with a step $\Delta\gamma=0.001$, and the sum of squares of residuals is calculated. The best fit corresponds to the minimal sum. 

Let us highlight that although the spectral statistics is universal in the extreme cases ($\gamma=0$ and $\gamma=1$), the intermediate spectral statistics may depend on the details of the Hamiltonian~\cite{Sierant_2019}. Thus, the Brody distribution is not the only possible choice for the intermediate situations (see also the Serbyn-Moore model with power-law interactions between levels, the $\beta$-Gaussian ensemble with fractional $\beta$, the Rosenzweig-Porten ensemble with a standard deviation as a free parameter, et cetera~\cite{Atas_2013b,Sierant_2017b,Bui_2019,Sierant_2019,De_2021}). Unfortunately, the low-energy mixtures of few fermions develop a noise on the top of the histograms of level spacings (and other measures of quantum chaos), what makes it difficult or even impossible to fully determine the character of intermediate statistics. We have verified, however, that the Brody distribution satisfactorily models the histogram of levels spacings in our system.

Let us mention that, in order to reduce the noise, we average $\left<\tilde{r}\right>_\text{n}$, $P\left(\tilde{r}\right)$ and ${\cal P} \left(\delta\right)$ over the eight nearest interaction strengths separated by $\Delta g=0.025$.

\begin{figure}
\includegraphics[width=\columnwidth]{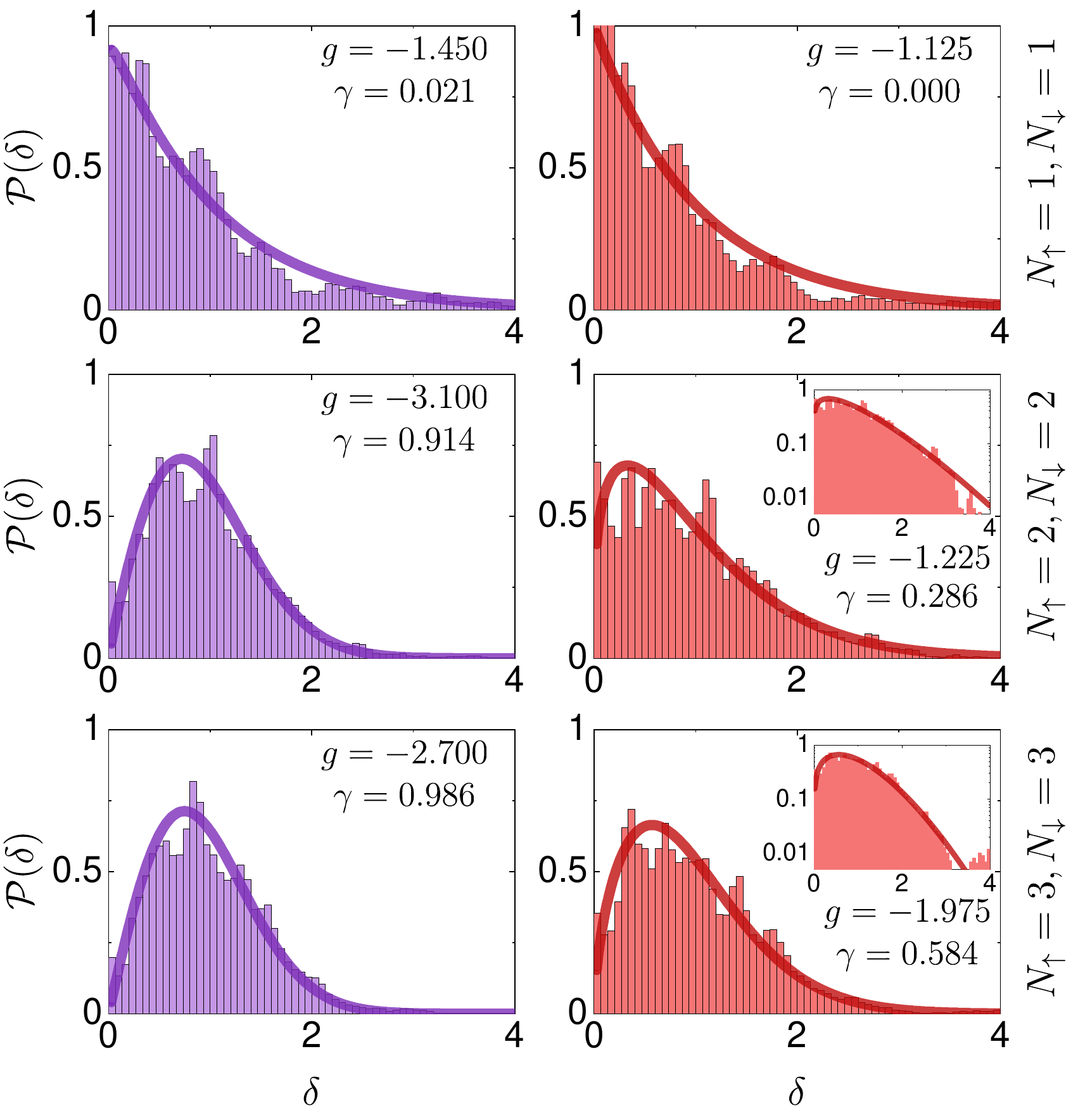}
\vspace{-0.6cm}
\caption{The histogram of spacings for $\mathcal{N}=150$ and the same interaction strengths as in Fig.~\ref{fig1}. The Brody distribution is marked with a solid line, while the Brody parameter $\gamma$ is indicated in the legend. The inset shows the histogram and distribution from the main panel in a logarithmic scale on the vertical axis.}
\label{fig1a}
\end{figure}
\section{Balanced mixtures}\label{sec::bal}
We first consider the balanced mixtures $N_\uparrow=N_\downarrow=N/2$ with $N=2,4$ and $6$ particles. Numerical results are collected in Fig.~\ref{fig1} and \ref{fig1a}. In the minimal example of two fermions $N_\uparrow=N_\downarrow=1$ (first row in Fig.~\ref{fig1}), the ratio $\left<\tilde{r}\right>_\text{n}$ differs from zero in an extremely narrow range of interaction strengths around $g\approx -2$ when $\mathcal{N}=50$. However, its value increases to $\left<\tilde{r}\right>_\text{n}\approx 0.5$ for $|g|\gtrsim\pm 1.4$ when the number of levels is increased to $\mathcal{N}\ge 150$. Simultaneously, the histogram of ratios $P\left(\tilde{r}\right)$ for the minimal $\left<\tilde{r}\right>_\text{n}=0$ follows the distribution for uncorrelated levels $P_\text{P}\left(\tilde{r}\right)$. It is almost unaffected by the increasing $\mathcal{N}$, except for the trivial smoothing of the noise (not shown). On the other hand, the histogram of ratios $P\left(\tilde{r}\right)$ for the maximal $\left<\tilde{r}\right>_\text{n}\approx 0.6$ does not resemble the distribution for random matrices $P_\text{GOE}\left(\tilde{r}\right)$. It has a pronounced maximum near $\left<\tilde{r}\right>_\text{n}=0$, and more of a picked-fence structure. We checked that the latter features are robust against the increase of $\mathcal{N}$. These results agree with the complementary approach to unveil correlations between energy levels, which is based on the histogram of spacings ${\cal P}\left(\delta\right)$ (first row of Fig.~\ref{fig1a}). It is clear that there are no correlation between energy levels. Recall that ${\cal P}\left(\delta\right)$ is somewhat sensitive to the unfolding method. Nevertheless, we have observed no level repulsion in all unfolding methods, so the mentioned sensitivity does not affect the overall conclusion about the minimal example of two fermions, {\it i.e.}, that it lacks typical features of quantum chaotic systems for all interaction strengths $g$. Throughout the paper, we do not discuss the weak interactions limit ({\it i.e.}, the close vicinity of $g=0$), in which massive degeneracies are observed, and the histograms of ratios $P\left(\tilde{r}\right)$ are not satisfactorily modelled by neither $P_\text{GOE}\left(\tilde{r}\right)$ nor $P_\text{P}\left(\tilde{r}\right)$.

When a particle is added to each component and $N_\uparrow=N_\downarrow=2$, the ratio $\left<\tilde{r}\right>_\text{n}$ becomes non-zero in a wide range of interaction strengths $g$, even for $\mathcal{N}=50$ (second row in Fig.~\ref{fig1}). Note the behaviour of alternating large and small values of $\left<\tilde{r}\right>_\text{n}$. The latter is somehow surprising, since it means that for some integrability-breaking perturbations $g$, a further increase of $g$ results in a suppression of quantum-chaotic features. However, this behaviour is lost when the number of levels is increased to $\mathcal{N} \ge 150$. The histograms of ratios $P\left(\tilde{r}\right)$ for the local minima of $\left<\tilde{r}\right>_\text{n}$ in the region of moderate interactions $g\in\left(-3,-1\right)$ are characterized by a Poisson-like tail for all $\mathcal{N}$. This signals that correlations between energy levels are highly local and weak. Therefore, the mixture is pseudo-integrable rather than quantum chaotic. On the contrary, the histograms of ratios $P\left(\tilde{r}\right)$ for the local maxima of $\left<\tilde{r}\right>_\text{n}$ follow the Wigner-Dyson distribution $P_\text{GOE}\left(\tilde{r}\right)$ for all $\mathcal{N}$. The analysis of correlation between energy levels based on the histograms of spacings ${\cal P}\left(\delta\right)$ (second row in Fig.~\ref{fig1a}) is in a full agreement with these predictions. For example, the relative difference between two measures of chaoticity, {\it i.e.}, the Brody parameter $\gamma$ and the ratio $\left<\tilde{r}\right>_\text{n}$, remains in the range  $0.02 - 0.12$. Therefore, we  claim the coexistence of pseudo-integrable and quantum-chaotic features in the mixture of four fermions for different interaction strengths $g$.

Finally, we address the largest studied example of six fermions $N_\uparrow=N_\downarrow=3$ (third row in Fig.~\ref{fig1}). The ratio $\left<\tilde{r}\right>_\text{n}$ is characterized by an almost monotonic growth followed by a saturation $\left<\tilde{r}\right>_\text{n}\approx 1$ for $g\gtrsim 2.7$. The histograms of ratios $P\left(\tilde{r}\right)$ for the small number of local minima of $\left<\tilde{r}\right>_\text{n}$ in the region of moderate interactions $g\in\left(-3,-1\right)$ vanish for $\tilde{r}\rightarrow 0$. Furthermore, $\left<\tilde{r}\right>_\text{n}$ increases and $P\left(\tilde{r}\right)$ approaches $P_\text{GOE}\left(\tilde{r}\right)$ with $\mathcal{N}$ (not shown). This indicates that correlations between energy levels become stronger for higher energies. Simultaneously, the histograms of ratios $P\left(\tilde{r}\right)$ for the remaining interaction strengths $g$ follow the Wigner-Dyson distribution $P_\text{GOE}\left(\tilde{r}\right)$. It seems the mixture of six fermions $N_\uparrow=N_\downarrow=3$ is quantum chaotic in the entire range of moderate interactions $g\in\left(-3,-1\right)$, provided that its energy is not too low. The histograms of spacings $P\left(\delta\right)$ can be used to draw the same conclusions, although the relative difference between the Brody parameter $\gamma$ and the ratio $\left<\tilde{r}\right>_\text{n}$ can reach larger values of $0.02 - 0.22$ (third row in Fig.~\ref{fig1a}).

The statistics of level spacings is a conventional measure of quantum chaos. However, it only recognizes correlations between the nearest energy levels. To confirm our findings with a complementary method, we calculate the averaged spectral form factor \cite{Cotler_2017,Liu_2018,Chen_2018}
\begin{equation}
\label{eqSFF}
K(t)=\left<\sum_{i,j} \exp\left(-i(E_i-E_j)t\right)\right>.
\end{equation}
The spectral form factor is not self-averaging, {\it i.e.}, the universal properties of quantum-chaotic systems are revealed after a moving average in a logarithmic time  $\left[\log t - \delta, \log t + \delta\right]$ is performed~\cite{Prange_1997,Bertini_2018}. The latter is marked as $\left<...\right>$ in Eq.~$\left(\ref{eqSFF}\right)$, and it is preceeded by the averaging over two symmetry sectors and eight interaction strengths. We fix $\delta=0.025$, and we calculate $K(t)$ out of $\mathcal{N}=150$ energy levels. Note that we omit the spectral unfolding of the many-body spectrum, since the secular part of the density has a non-trivial effect only on the early-time part of $K(t)$, see Ref.~\cite{G_2018} and \cite{Prakash_2021}.

It has been established that $K(t)$ of quantum-chaotic systems develops a correlation hole (a linear ramp) in the moderate time, which is not observed in the case of integrable systems~\cite{G_2018,Suntajs_2020,Suntajs_2021,Prakash_2021,Fogarty_2021}. Generally, the time at which the averaged spectral form factor enters a linear ramp, {\it i.e.}, the Thouless time, sets the time scale at which the dynamics of a system becomes universal. It is related to the inverse of the energy scale at which the correlations between energy levels predicted by the random matrix theory are no longer present. Interestingly, $K(t)$ of quantum-chaotic quadratic systems (with the Wigner-Dyson statistics in the single-particle sector but the Poisson statistics in the many-body one~\cite{Lydzba_2021a,Lydzba_2021b}) develops an exponential ramp~\cite{Liao_2020,Winer_2020}. The latter originates from correlations between distant many-body levels, whose separation is comparable to the mean level spacing between single-particle levels.

\begin{figure}
\includegraphics[width=\columnwidth]{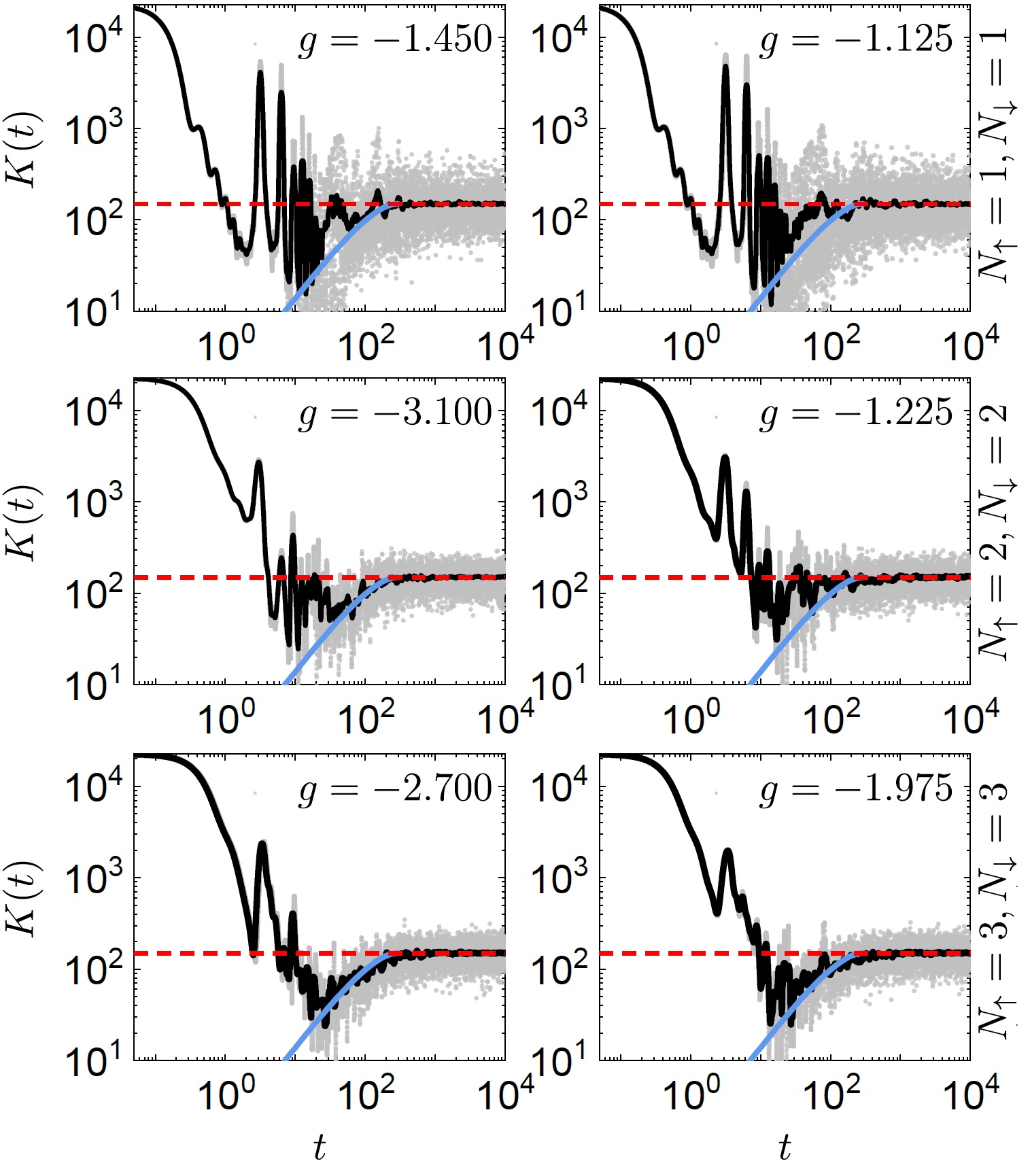}
\vspace{-0.6cm}
\caption{The spectral form factor for balanced mixtures with $N_\uparrow=N_\downarrow=1,2,3$ and $\mathcal{N}=150$. The same interaction strengths as in Fig.~\ref{fig1} are considered. The black line corresponds to the moving average of the spectral form factor in the logarithmic time window $\left[\log t - \delta, \log t + \delta\right]$ with $\delta=0.025$, while the grey points mark the unaveraged data. The blue line marks the well-known moderate time results for the Gaussian Orthogonal Ensemble $K_\text{GOE}\propto \alpha t-\frac{\alpha t}{2}\ln\left(1+\alpha t\right)$~\cite{Liu_2018}, which is rescaled so that the mean level spacing is consistent with our spectrum~\cite{Prakash_2021}. The red dashed line is the long time result, which is the same for quantum chaotic and integrable systems. }
\label{fig1b}
\end{figure}

Spectral form factors for the same setups as in Fig.~\ref{fig1} are presented in Fig.~\ref{fig1b}. They confirm our previous findings. In the minimal scenario of two fermions $N_\uparrow=N_\downarrow=1$, the correlation hole is absent in the entire range of interactions $g\in\left[-4,2\right]$ (first column of Fig.~\ref{fig1b}), When the particle numbers are incerased to $N_\uparrow=N_\downarrow=2$, the pseudo-integrable regions of $g$ coexist with the quantum-chaotic regions of $g$ (second row of Fig.~\ref{fig1b}). Nevertheless, the spectral form factor enters the linear ramp at a relatively late time in the latter region, signalling that the energy scale of correlations or the fraction of correlated energy levels is relatively small. If the considered part of the many-body spectrum has a sharp spectral edge, the time of occurrence of the linear ramp can differ from the Thouless time~\cite{G_2018}. The latter is unavoidably present in our low energy mixtures of few fermions. Since we cannot employ the Gaussian filtering introduced in Ref.~\cite{G_2018}, we discuss the qualitative rather than quantitative behaviour of the linear ramp. Finally, in the largest system with $N_\uparrow=N_\downarrow=3$, the correlation hole is established in the entire range of interactions  $g\in\left[-4,2\right]$ (third row of Fig.~\ref{fig1b}).

\begin{figure}[t]
\includegraphics[width=\columnwidth]{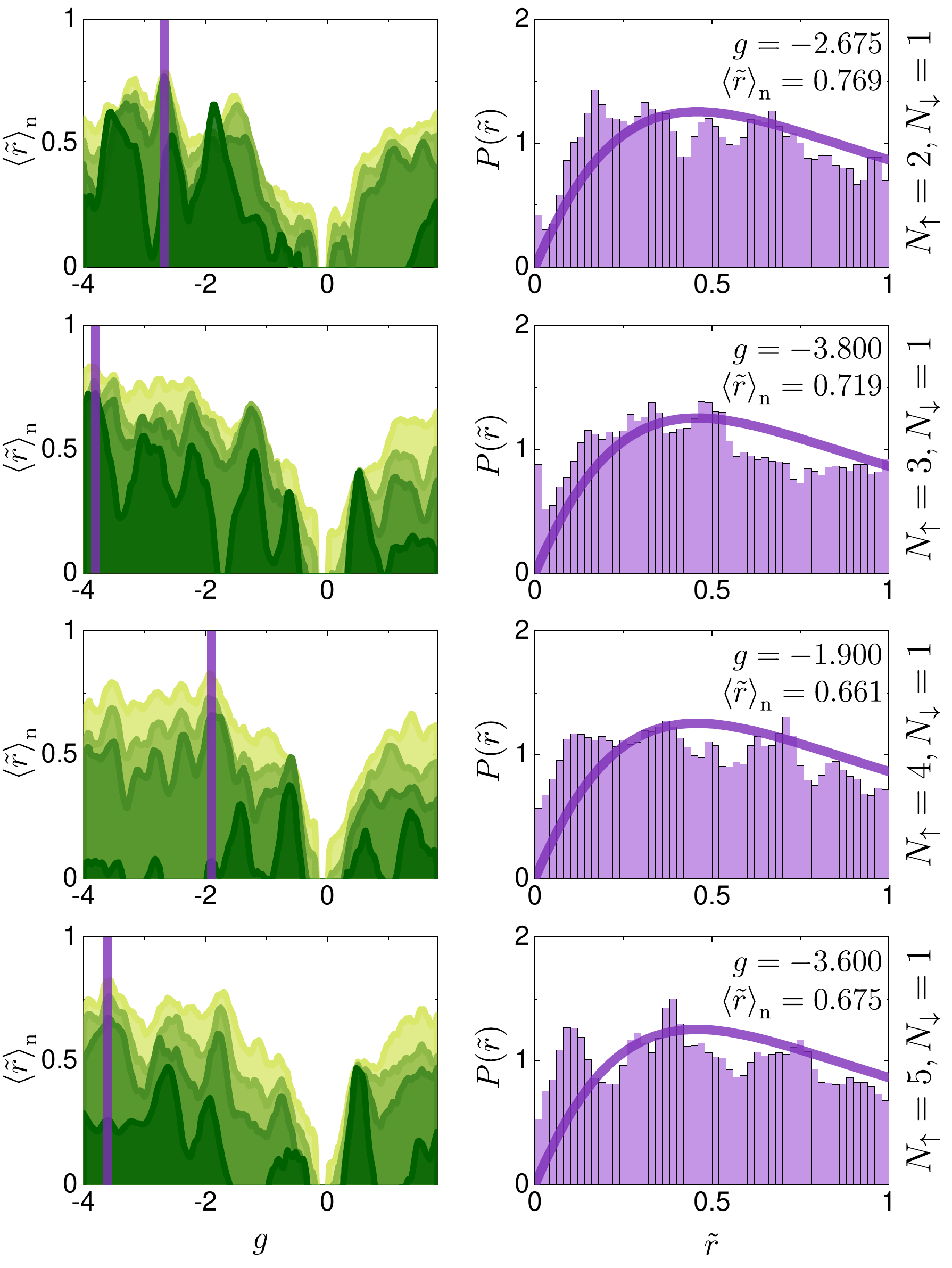}
\vspace{-0.2cm}
\caption{Spectral properties of imbalanced mixtures with $N_\downarrow=1$ and $N_\uparrow =2,3,4,5$ particles. The rescaled ratio $\left<\tilde{r}\right>_\text{n}$ is presented in the first column. We consider $\mathcal{N}=50,150,250$ and $500$, and numerical results for a larger $\mathcal{N}$ are marked by a brighter green colour. The histogram $P\left(\tilde{r}\right)$ for the maximal $\left<\tilde{r}\right>_\text{n}$ and $\mathcal{N}=150$ is presented in the second column [together with the distribution from Eq.~$\left(\ref{eqGOE}\right)$]. The interaction strength $g$ corresponding to the maximal $\left<\tilde{r}\right>_\text{n}$ is marked by a purple solid line in the first column.}
\label{fig2}
\end{figure}
\section{Imbalanced mixtures}\label{sec::imb}
Let us now deal with the scenario in which the number of particles is increased in one component $N_\uparrow=2,3,4$ and $5$, while it is kept fixed in the other component $N_\downarrow=1$. Numerical results are gathered in Fig.~\ref{fig2}. When $N_\uparrow=1$ is changed to $N_\uparrow=2$, the range of interactions~$g$ with the ratio $\left<\tilde{r}\right>_\text{n}\approx 0$ clearly decreases. Moreover, the alternating minima and maxima of $\left<\tilde{r}\right>_\text{n}$ are formed for $\mathcal{N}=50$. They are blurred for $\mathcal{N}\ge 150$. As apparent from the histogram of ratios $P\left(\tilde{r}\right)$ for the maximal $\left<\tilde{r}\right>_\text{n}$, the correlations between energy levels and, so, the chaoticity are stronger when compared to the minimal example of two fermions $N_\uparrow=1$. Moreover, the maximal $\left<\tilde{r}\right>_\text{n}$ slowly increases with $\mathcal{N}$. It should be emphasized that even for the largest number of levels $\mathcal{N}=500$, the histogram of ratios $P\left(\tilde{r}\right)$ is different than the distribution for random matrices $P_\text{GOE}\left(\tilde{r}\right)$.  It is worth mentioning that some (not fully developed) signatures of quantum chaos were recently observed in a one-dimensional system of three particles, but in a different potential and in the high energy limit~\cite{Fogarty_2021}.

If particles are further added to the component ($N_\uparrow=3,4,5$), the chaoticity becomes suppressed. The maximal ratio $\left<\tilde{r}\right>_\text{n}$ decreases with $N_\uparrow$, but slowly increases with $\mathcal{N}$. The histogram $P\left(\tilde{r}\right)$ in the limit $\tilde{r}\rightarrow 0$ becomes more pronounced and, so, the level repulsion becomes less effective upon increasing $N_\uparrow$. At the same time, a slow opposite trend is observed upon increasing $\mathcal{N}$ (not shown). Note that sharp peaks appear on the top of $P\left(\tilde{r}\right)$ for $N_\uparrow\ge 4$ that signal the restoration of the picket-fence structure.

As a result, we have determined that keeping the considered mixture of fermions close to the balanced scenario enhances its chaoticity, and suggests that one should not expect robust quantum-chaotic correlations in the impurity limit ($N_\uparrow\rightarrow\infty$). Note that this outcome lies in contrast to the results for an impurity embedded in a bosonic bath restricted to two modes of a double-well potential~\cite{Chen_2021}, and exposes the importance of quantum statistics for unravelling the presence of quantum chaos in ultracold ensembles.

\begin{figure}[t]
\includegraphics[width=\columnwidth]{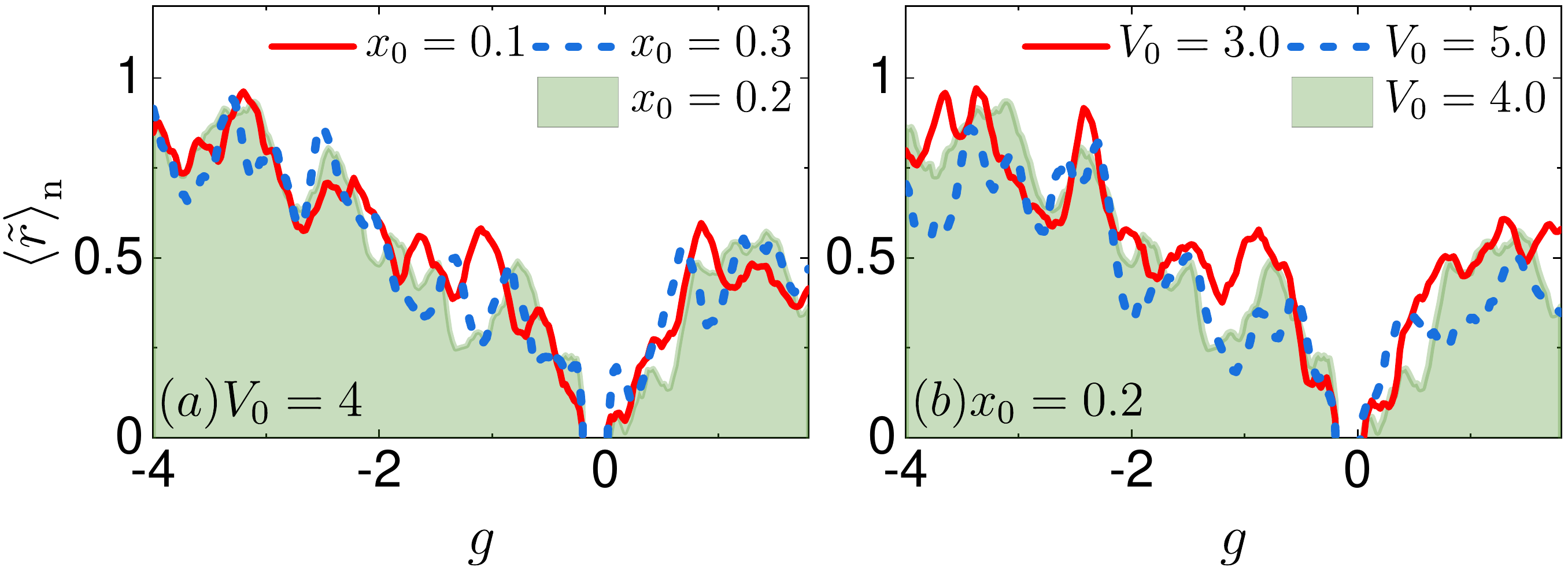}
\vspace{-0.2cm}
\caption{ The rescaled ratio $\left<\tilde{r}\right>_\text{n}$ for a balanced mixture with $N_\uparrow=N_\downarrow=2$ and $\mathcal{N}=150$. Results for a fixed height $V_0=4.0$ and different widths $x_0$ are presented in (a), while for a fixed width $x_0=0.2$ and different heights $V_0$ are presented in (b). The green area serves as a reference result with $V_0=4.0$ and $x_0=0.2$. It has been already presented in Fig.~\ref{fig1}.}
\label{fig3}
\end{figure}
\section{Role of confining potential} \label{sec::rol}
The shape of a confining potential usually has an effect on the correlations between energy levels. Therefore, we checked how the spectral statistics and, so, the chaoticity of a two-component mixture of few fermions changes when the parameters $V_0$ and $x_0$ of a double-well potential are varied. The ratio $\left<\tilde{r}\right>_\text{n}$ determined for the system with $N_\uparrow=N_\downarrow=2$ particles and $\mathcal{N}=150$ levels is presented in Fig.~\ref{fig3}. Despite some quantitative differences, all plots are qualitatively similar. For example, the global trend of $\left<\tilde{r}\right>_\text{n}$ is independent of the barrier height $V_0$ (the right panel of Fig.~\ref{fig3}) as well as the barrier width $x_0$ (the left panel of Fig.~\ref{fig3}). Moreover, local extrema are formed near similar interaction strengths $g$. As a result, the discussion about the emergence of quantum chaos is hardly affected by minor modifications of the confining potential.

\section{Concluding remarks} \label{sec::con}
In the heart of the eigenstate thermalization hypothesis (ETH) lies the assumption that individual energy eigenstates of quantum-chaotic systems obey the laws of statistical mechanics, provided that their energies are not too low~\cite{Deutsch_2018,Srednicki_1994}. The main consequence of the ETH is that a quantum-chaotic system can thermalize when driven out of equilibrium, even if it is isolated from the external environment~\cite{Rigol_2007,Rigol_2009,Rigol_2011,Khatami_2012,Caux_2013,Reimann_2016,Lydzba_2021}. This has been confirmed in recent ultracold atom experiments~\cite{Trotzky_2012,Kaufman_2016,Clos_2016,Tang_2018}. The eigenstate thermalization hypothesis is closely related to the random matrix theory (RMT)~\cite{Schonle_2021,Jansen_2019,D_Alessio_2016}. The existence or absence of the RMT-like correlations in the many-body spectrum has been investigated in many systems, even mesoscopic ones~\cite{Fogarty_2021,Mirkin_2021,Santos_2010}. However, rarely in the low-energy limit. Note that the measurement of the spectral statistics is tricky, but has been successfully performed using slow neutron resonances of heavy nuclei and poton resonances of light nuclei~\cite{Bohigas_1983,Mitchell_1991} (see also the measurement based on acoustic resonances in quartz blocks~\cite{Ellegaard_1996}).

As pointed out in Ref.~\cite{Mirkin_2021}, the most recent experiments often consist of working with few ultracold particles and executing controlled operations on some of them. The maximum degree of control achievable in an arbitrary subsystem is limited by the ability of a non-controlled part of the system to scramble information or act as an internal environment. The possibility of an internal equilibration after a perturbation is a property of quantum chaotic systems. In order to understand and efficiently minimize errors in experimental implementations of controlled operations, it is necessary to understand whether and how quantum chaos is manifested in mesoscopic systems. In the low-energy limit as well.

In this paper, we considered a two-component mixture of few fermions, which interacted via contact interactions and were confined in a one-dimensional double-well potential. We studied spectral measures of quantum chaos, {\it i.e.}, the averaged rescaled ratio $\left<\tilde{r}\right>_\text{n}$, the distribution of the ratio $P\left(\tilde{r}\right)$, the distribution of the spacing ${\cal P}\left(\delta\right)$ as well as the spectral form factor $K(t)$.
These studies were restricted to $\mathcal{N}=50,150,250$ and $500$ levels in the low-energy tail of the many-body spectrum. We observed some signatures of quantum chaos, {\it i.e.}, the level repulsion, already in the system with three particles $N_\uparrow=2$ and $N_\downarrow=1$. Generally, these signatures become weaker when the number of particles is increased in the imbalanced scenario, {\it i.e.}, when fermions are added to one component $N_\uparrow$, while the other component is fixed $N_\downarrow=1$. The latter suggests that the RMT-like correlations are not expected in the impurity limit $N_\uparrow\rightarrow\infty$. On the contrary, these signatures become more pronounced when the number of particles is increased in the balanced scenario $N_\uparrow=N_\downarrow$. In the system with four particles $N_\uparrow=N_\downarrow=2$, the regimes of interactions with the RMT-like correlations between the nearest energy levels coexist with the regimes of interactions with almost no correlations between the nearest energy levels. Simultaneously, in the system with six particles $N_\uparrow=N_\downarrow=3$, signatures of quantum chaos are witnessed in the entire range of interactions $g\in[-4,2]$.

\acknowledgements
The authors thank L. Vidmar and M. Rigol for fruitful discussions concerning quantum-chaotic systems and their spectral properties. TS acknowledges hospitality from UPV in Val\`encia.

This work has been supported by the Slovenian Research Agency (ARRS), Research core fundings Grant No.~P1-0044 (P\L), and the (Polish) National Science Centre Grant No.~2016/22/E/ST2/00555 (TS).

\appendix 

\section{Energy cut-off method}\label{Zalacznik} 
\begin{figure}[t]
\includegraphics[width=\columnwidth]{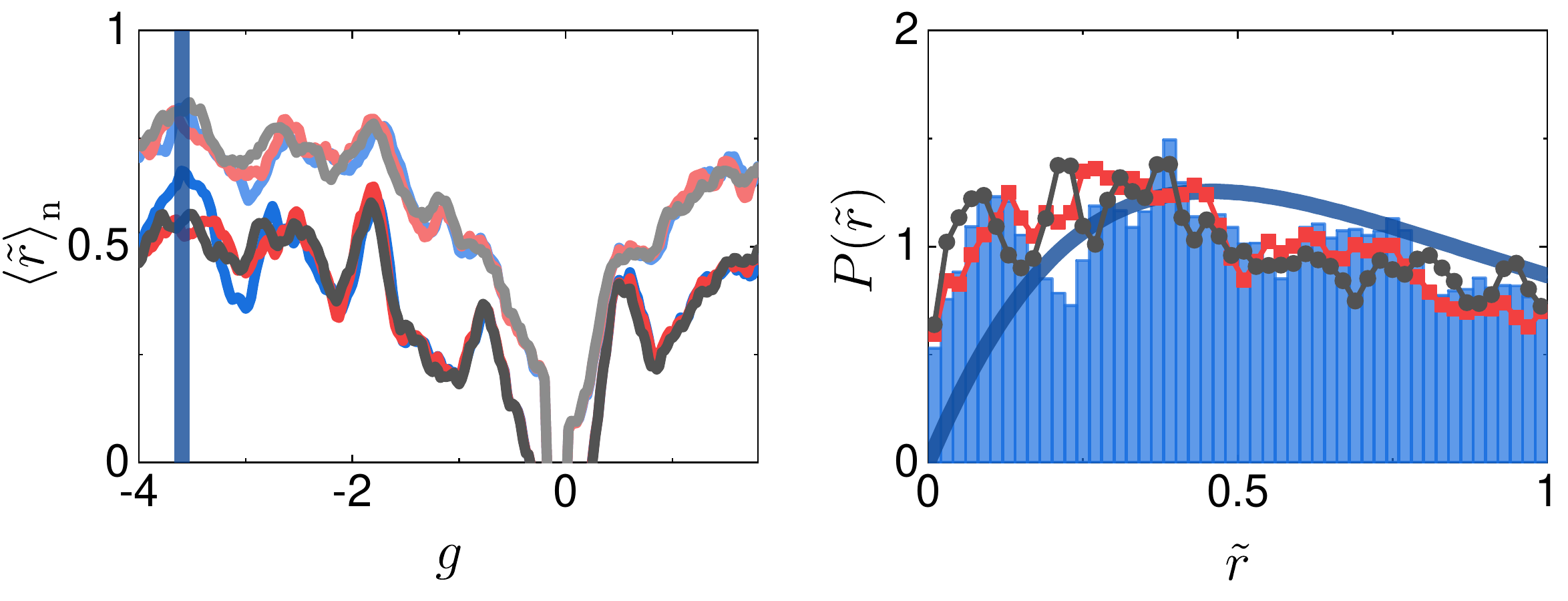}
\vspace{-0.2cm}
\caption{The rescaled ratio $\left<\tilde{r}\right>_\text{n}$ in the largest imbalanced system $N_\uparrow=5$ and $N_\downarrow=1$ (left panel). Three cut-off energies $E_\text{max}=35$ (grey lines), $40$ (red lines) and $45$ (blue lines) are considered. Moreover, numerical results for two numbers of levels $\mathcal{N}=150$ (dark lines) and $500$ (bright lines) are presented. The distribution of ratios $P\left(\tilde{r}\right)$ for the interaction strength $g=-3.6$ and $\mathcal{N}=150$ (right panel). The interaction strength $g=-3.6$ is marked by the vertical line in the left panel. The (red) gray line-scatter plot corresponds to $E_\text{max}=35$($40$), while the bar plot corresponds to $E_\text{max}=45$. The blue curve marks the Wigner-Dyson distribution. }
\label{fig4}
\end{figure}
The energy cut-off method generates all Fock states with the non-interacting energy satisfying the inequality $\langle\hat{H}_{0}\rangle \leq E_\text{max}$.The simplest way to describe the procedure is to represent Fock states in the first quantization notation, {\it i.e.}, with two algebraic vectors $\vec{O}_\uparrow = (O_{\uparrow 1},\ldots,O_{\uparrow N_\uparrow})$ and $\vec{O}_\downarrow=(O_{\downarrow 1},\ldots,O_{\downarrow N_\downarrow})$ that gather indices of occupied single-particle levels. By the definition, the consecutive numbers in these vectors are in the ascending order. In this notation, the non-interacting ground state is represented by $\vec{O}_\uparrow=(0,1,\ldots,N_\uparrow-1)$ and $\vec{O}_\downarrow=(0,1,\ldots,N_\downarrow-1)$. To generate all non-interacting excited states with the energy bounded by $E_\text{max}$, one performs the following steps. 
\begin{enumerate}
\item We take the ground state as a temporal state $\mathtt S$.
\item We set the index $I$ to $N_\downarrow$.
\item We determine whether the energy of the temporal state $\mathtt S$ is smaller than $E_\text{max}$. If the bound is not violated, we go to 5.
\item We decrease the index, $I\rightarrow I-1$. If the index is smaller than 1 the algorithm is finished. Otherwise we go to 6.
\item We accept the temporal state $\mathtt S$.
\item We change the temporal state $\mathtt S$ along the following rule. We excite the $I$-th particle in the $\downarrow$-component, $O_{\downarrow I}=O_{\downarrow I}+1$, and reset the state of all particles with higher indices, {\it i.e.}, for all $k>I$ we set $O_{\downarrow k}=O_{\downarrow k-1}+1$.  
\item We go back to 3.
\end{enumerate}
The described procedure allows us to determine all $O_\downarrow$, for which the first component is in the ground state configuration and $\langle\hat{H}_0\rangle \le E_\text{max}$ Subsequently, the protocol is repeated for other possible configurations of the first component, which can be determined in the same spirit as all possible configurations of the second component. More details on the energy cut-off method can be found in~\cite{Chrostowski_2019}.

Before we finish this section, we demonstrate how two measures of quantum chaos studied in the paper, {\it i.e.}, the rescaled ratio $\langle\tilde{r}\rangle_\text{n}$ and the distribution $P\left(\tilde{r}\right)$, change with $E_\text{max}$. Numerical results for the largest imbalanced system $N_\uparrow=5$ and $N_\downarrow=1$ are presented in Fig.~\ref{fig4}. We consider three cut-off energies $E_\text{max}=35$ (grey lines), $40$ (red lines) and $45$ (blue lines). The rescaled ratio $\left<\tilde{r}\right>_\text{n}$ is presented for two numbers of levels $\mathcal{N}=150$ (dark lines) and $\mathcal{N}=500$ (bright lines), see the left panel of Fig.~\ref{fig4}. Minor differences between $\left<\tilde{r}\right>_\text{n}$ obtained with different $E_\text{max}$ are revealed for $g<-3$. Nevertheless, the general shape of the density $P\left(\tilde{r}\right)$ remains independent of $E_\text{max}$, see the right panel of Fig.~\ref{fig4}.

\section{SU(2) symmetry}\label{Zalacznik2} 
\begin{figure*}[t]
\includegraphics[width=\textwidth]{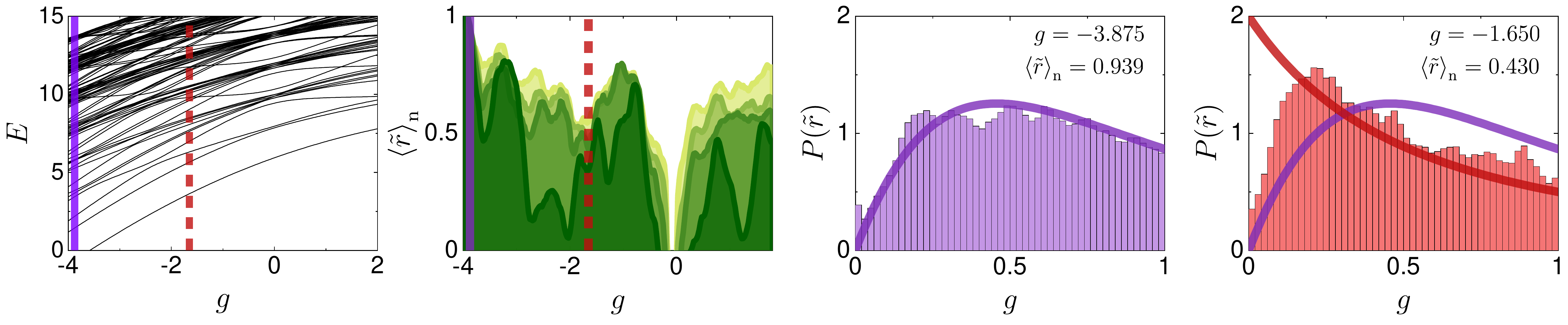}
\vspace{-0.2cm}
\caption{Spectral properties of a balanced mixture with $N_\uparrow=N_\downarrow=2$ particles in the lowest spin sector $S=0$. The low-energy part of the many-body spectrum is displayed in the first column. The rescaled ratio $\left<\tilde{r}\right>_\text{n}$ is presented in the second column. We consider $\mathcal{N}=50,150,250$ and $500$, and numerical results for a  larger $\mathcal{N}$ are marked by a brighter green colour. The histogram $P\left(\tilde{r}\right)$ for a maximum of $\left<\tilde{r}\right>_\text{n}$ and $\mathcal{N}=150$ is presented in the third column [together with the distribution from Eq.~$\left(\ref{eqGOE}\right)$]. Simultaneously, the histogram $P\left(\tilde{r}\right)$ for a minimum of $\left<\tilde{r}\right>_\text{n}$ and $\mathcal{N}=150$ is presented in the fourth column [together with the distribution from Eq.~$\left(\ref{eqP}\right)$]. The interaction strength $g$ corresponding to a maximum (minimum) of $\left<\tilde{r}\right>_\text{n}$ is marked by a purple solid (red dashed) line in the first two columns. }
\label{fig5}
\end{figure*}

The Hamiltonian~\eqref{eqH} is SU($2$) invariant, {\it i.e.}, it commutes with $\hat{S}_{z}$ and $\hat{S}^2$ operators. In the following, we express the projection and the squared magnitude of the total spin in units of $\hbar$ and $\hbar^2$, respectively. Commutation with spin projection is insignificant, since $\hat{S}_{z}=\frac{1}{2}\left(N_\uparrow - N_\downarrow\right)$. Thus, its eigenvalue $S_z$ is fixed in the subspaces with well-defined particle numbers $N_\uparrow$ and $N_\downarrow$. On the other hand, by taking advantage of the decomposition of the field operator, one finds that $\hat{S}^2 = \sum_{ij}\hat{a}^\dagger_{\uparrow i}\hat{a}_{\downarrow i}\hat{a}^\dagger_{\downarrow j}\hat{a}_{\uparrow j}+\hat{S}_{z}(\hat{S}_{z}-1)$. It means that the energy spectrum comprises uncorrelated parts, each labeled by a different eigenvalue $S(S+1)$ where $S\geq S_z$. When the number of particles $N=N_\uparrow + N_\downarrow$ is even, the magnitude of the total spin $S$ is an integer with $S\in\left\{0,1,...,N/2\right\}$. Simultaneously, when the number of particles $N=N_\uparrow + N_\downarrow$ is odd, the magnitude of the total spin $S$ is a half-integer with $S\in\left\{1/2,3/2,...,N/2\right\}$. Therefore, there are always two symmetry sectors in the imbalanced mixtures with $N_\uparrow \le 5$ and $N_\downarrow=1$, while two, three and four symmetry sectors in the balanced mixtures with $N_\uparrow=N_\downarrow=1,2$ and $3$ particles, respectively.

For all the systems considered in the main part of the paper, we have calculated the expectation values of $\hat{S}^2$ in the many-body states with the lowest energies. We have found out that only spin sectors with $S<N/2$ have significant contributions to the tail of the spectrum (there are at most few states with $S=N/2$). Generally, the lower the magnitude of the total spin $S$, the higher the population of many-body states. This means that only one spin sector with $S=S_z$ is active in the imbalanced mixtures of fermions with $N_\uparrow \le 5$ and $N_\downarrow=1$. On the other hand, it seems that up to two spin sectors are active in the balanced mixtures of fermions with $N_\uparrow=N_\downarrow=1,2$ and $3$.

In order to determine how the SU($2$) symmetry modifies the spectral statistics, we have diagonalized the following matrix $\hat{M}=\hat{H}+\beta \hat{S}^2$. The parameter $\beta$ is selected so that the expectation value of $\hat{S}^2$ is fixed in the tail of the spectrum (i.e., $S=S_z$). Subsequently, we have calculated expectation values of $\hat{H}$ and ordered them in the ascending order. Finally, we have recalculated the rescaled ratio $\left<\tilde{r}\right>_\text{n}$ and the whole distribution of ratios $P\left(\tilde{r}\right)$. As expected, the latter measures of quantum chaos are indistinguishable from the ones presented in the main part from the paper for the imbalanced mixtures of fermions. Although they do change at the quantitative level, they remain the same at the qualitative level for the balanced mixtures of fermions. Numerical results for $N_\uparrow=N_\downarrow=2$ are presented in Fig.~\ref{fig5}.

\bibliographystyle{style}
\bibliography{references}


\end{document}